# Screening of nuclear reactions in the Sun and solar neutrinos

B. Ricci$^{(1,2)}$, S. Degl'Innocenti$^{(2,3)}$ and G. Fiorentini$^{(2,3)}$
$^{(1)}$*Scuola di Dottorato dell'Università di Padova, I-35100 Padova,*
$^{(2)}$*Istituto Nazionale di Fisica Nucleare, Sezione di Ferrara, I-44100 Ferrara,*
$^{(3)}$*Dipartimento di Fisica dell'Università di Ferrara, I-44100 Ferrara.*
(November 22, 1994)

## Abstract

We quantitatively determine the effect and the uncertainty on solar neutrino production arising from the screening process. We present predictions for the solar neutrino fluxes and signals obtained with different screening models available in the literature and by using our stellar evolution code. We explain these numerical results in terms of simple laws relating the screening factors with the neutrino fluxes. Futhermore we explore a wider range of models for screening, obtained from the Mitler model by introducing and varying two phenomenological parameters, taking into account effects not included in the Mitler prescription. Screening implies, with respect to a no-screening case, a central temperat reduction of 0.5%, a 2% (8%) increase of Beryllium (Boron)-neutrino flux and a 2% (12%) increase of the Gallium (Chlorine) signal. We also find that uncertainties due to the screening effect ar at the level of 1% for the predicted Beryllium-neutrino flux and Gallium signal, not exceeding 3% for the Boron-neutrino flux and the Chlorine signal.

96.60.Kx

# I. INTRODUCTION

The solar neutrino problem is so important that any aspect of solar, plasma and nuclear physics pertinent to it has to be deeply investigated before definitive conclusions can be drawn, for recent reviews see for example [1–3].

In this respect, screening of the charges of the reacting nuclei due to free charges in the solar plasma is of some interest. The study of screened nuclear reaction rates was started with the pioneer work of Salpeter [4]; it has been investigated by several authors, see for example [5–8], and recently reviewed in [9]. In the Sun, the screening effect is relatively small but the situation is not completely clear to us, as different calculations yield relatively different nuclear reaction rates.

The electron screening of nuclear reactions in the laboratory has been recently investigated experimentally and theoretically for atomic and nuclear targets, see for example [10,11]. The effect has been observed and it generally comes out to be larger than theoretically predicted, even in the Born-Oppenheimer approximation (note however that experimental errors are large at the moment). Although this argument refers to a different context, it provides an incentive for additional investigation about screening of nuclear reactions in the solar plasma.

The aim of this paper is to quantitatively determine the effect and the uncertainty on solar neutrino production arising from the screening process, and with this in mind:
i) we present results for the solar neutrino fluxes and predicted signals obtained by referring to different screening prescriptions available in the literature and by using the FRANEC evolution code, for details about FRANEC code see [12].
ii) We explain the numerical results, in terms of simple laws relating the screening (enhancement) factors with the neutrino fluxes.
iii) We explore a wider range of models for screening obtained from the Mitler model, by introducing and varying two phenomenological parameters which can account for physical effects not included in the original Mitler treatment. We also perform a model independent analysis, where the enhancement factors for the $pp$ and the $He+He$ reactions are kept as free variables, and compare the prediction with solar neutrino experimental results.

## II. RESULTS OF SOLAR MODEL CALCULATIONS FOR DIFFERENT SCREENING PRESCRIPTIONS

We consider solar models based on five different assumptions:
i) Neglect completely any screening effect (NOS), i.e. nuclear reactions occur for bare ions with rate $\lambda_{\text{bare}}$.
ii) The weak screening approximation (WES) originally introduced by Salpeter [4] for a Debye plasma, where electron partial degeneracy is accounted. The reaction rate is now $\lambda = \lambda_{\text{bare}} f$ where the enhancement factor $f$ is given by

$$ln f^{\text{WES}} = Z_1 Z_2 e^2 \kappa / (kT) \qquad (1)$$

with $Z_{1,2}$ being the charges of reacting nuclei, T is the temperature and $\kappa$ is the inverse of the Debye radius, given by:

$$\kappa^2 = \frac{4\pi e^2}{kT}(\sum_i Z_i^2 n_i + n_e \theta_e) \qquad (2)$$

where $n_i$ is the number density of ions with charge $Z_i$, $n_e$ is the average electron density and $\theta_e$ is the electron degeneracy factor. In equation (2) the first term inside brackets correspond to ionic screening, whereas the latter one corresponds to electron screening. Note that electrons are essentially classical in the Sun, the degeneracy factor being $\theta_e = 0.93$ in the solar center and approaching the classical limit ($\theta_e = 1$) as moving outwards.

As well known, in the Sun the weak screening approximation can be justified (to some extent) for the $pp$-reaction, whereas the other nuclear reactions occur in the so-called intermediate screening regime.

iii) The Graboske et al. result [6] (GDGC): the enhancement factor is given by eq (1) when $lnf^{\text{WES}}$ is less than 0.1, while for larger values (up to 2) it is obtained by using general thermodynamic arguments and interpolation of Montecarlo calculations. Thus the enhancement factors are given now by:

$$lnf^{\text{GDGC}} = \begin{cases} lnf^{\text{WES}} & \text{for} \quad lnf^{\text{WES}} \leq 0.1 \\ 0.380 \frac{\langle z^{3b+1}\rangle}{\tilde{z}^{3b-2}\bar{z}^{2-2b}}[(Z_1+Z_2)^{1+b} - Z_1^{1+b} - Z_2^{1+b}]/(kT) & \text{for } 0.1 \leq lnf^{\text{WES}} \leq 2 \end{cases}, \qquad (3)$$

with $b = 0.860$, $\langle z^{3b+1}\rangle = \sum_i Z_1^{3b+1} X_i/A_i$, $\tilde{z} = \sum_i Z_i^2 X_i/A_i + \theta_e \bar{z}$, $\bar{z} = \sum_i Z_i X_i/A_i$, ($Z_i, X_i, A_i$ being respectively the atomic number, the mass number and the abundance in mass of the i-th component of solar plasma). In practice, in the solar interior the $pp$ reaction is calculated according to equation (3.1), all other reactions according to equation (3.2). As already remarked in ref [9], this prescription yields an unphysical discontinuity at the border between the weak and intermediate regimes.

iv) The Mitler result [8] (MIT), obtained with an analytical method which goes beyond the linearized approach and which correctly reproduces both the limits of weak and strong screening. Neglecting the small effects of a radial dependence in the effective potential, see [9], the enhancement factors are given now by:

$$lnf^{\text{MIT}} = -\frac{8}{5}\frac{(\pi e n_o)^2}{\kappa^5}[(\zeta_1+\zeta_2+1)^{5/3} - (\zeta_1+1)^{5/3} - (\zeta_2+1)^{5/3}]/(kT) \qquad (4)$$

where $\zeta_{1,2} = 3\kappa^3 Z_{1,2}/4\pi n_o$ and $n_o$ is the electron number density around the the interacting nuclei. Following Mitler we take it to be equal to the average electron density:

$$n_o = n_e \qquad (5)$$

v) The Carraro et al. result [7] (CSK): this takes into account that the reacting nuclei move faster than most of the plasma ions (the Gamow peak energy is generally larger than thermal energy), so the ion screening plays a smaller role under this condition. The dynamic response of the plasma is then calculated in the framework of the Debye theory. The resulting enhancement factors are thus expressed in terms of those of the weak screening:

$$lnf^{\text{CSK}} = C \; lnf^{\text{WES}} \quad . \qquad (6)$$

The correction factors C at the center of the Sun are $C_{p+p} = 0.76$, $C_{^3He+^3He} = 0.75$, $C_{^3He+^4He} = 0.76$, $C_{p+^7Be} = 0.80$, $C_{p+^{14}N} = 0.82$ and we assume they are constant along the solar profile.

The results of the corresponding solar models are shown in Table I where one notes the following features:
i) The largest differences arise between the NOS model and the WES models. The Boron neutrino flux can be varied at most by a 15% whereas the Chlorine signal is stable within 13% and the Gallium signal at the level of 3%.
ii) The GDGC model, which is extensively used in stellar evolution codes, yields values very close to the NOS model, the difference between the two being at the level of 1% for the Beryllium and Boron-neutrino fluxes, as well as for the Chlorine and Gallium signals.

In order to understand the role of screening effects it is useful to look at the enhancement factors along the solar profile, calculated by using the different prescription outlined above, for the reactions relevant to hydrogen burning in the Sun, see Fig. 1. We note that in all the models we are considering, there is no trace of a isotopic dependence (a part possibly for the CSK model) and concerning the CNO cycle we pay attention just to the slowest reaction $p + ^{14}N$.

As it is clear from Fig. 1, all the enhancement factors depend very weakly on the mass coordinate, at least as long as the energy production region ($M/M_\odot < 0.3$) is concerned. This is clear in the weak screening regime, since the dependence on the solar structure parameters (see equations [1] and [2]), is just of the form $\rho/T^3$ and this quantity, as well known, is approximately constant along the solar profile. The same holds in the strong screening regime, and thus the approximate constancy in the intermediate regime is not a surprise. For these reasons, in the following we will concentrate on the enhancement factors calculated at the solar center, see Table II.

As one sees from Fig.1 and Table II, the weak screening approximation, eq. (1), always yields the largest enhancement factors, as physically clear due to the fact that electrons and ions are assumed to be free and capable of following the reacting nuclei, and also the electron cloud is allowed to strongly condense around the nuclei (in the Debye limit the electron density becomes infinite at the nuclear site).

In the Mitler model, where electron density at nuclear site is fixed at $n_e$, the enhancement factor is smaller. The same holds for the CSK model where the limited mobility of ions and thus their partial screening capability is taken into account.

The GDGC enhancement factors are systematically smaller then the others (a part for the *pp* reactions where by definition they are equal to the WES prescription). In this respect it is clear that the neutrino fluxes and experimental signals calculated by using the GDGC prescription are the closest ones to the no-screening models. Similarly one understands the reason for the most marked difference being those between the WES solar model and NOS solar model.

A comparison with the results of ref. [9] (in particular see Table 3, Table 4 and Fig. 5) is interesting. Generally we agree with these authors but for a few points: i) we find a (small) difference, as screening is varied, for the predicted Beryllium-neutrino flux (see next section). ii) We generally have a higher Boron-neutrino flux, consistently with the different physical input parameters (S-factors, chemical composition, age) we are using in our evolution code (see Tables III and IV in ref. [16]), but the screening dependence is anyhow similar.

## III. ENHANCEMENT FACTORS AND NEUTRINO FLUXES

The influence of the screening effect on the neutrino production can be quantitatively understood in the approximation that the enhancement factors are constant in the energy production region. In this case the introduction of the enhancement factor $f_{i+j}$ for the reaction between nuclei i and j is equivalent to a changement of the zero-energy astrophysical S factors:

$$S_{i+j} \to S_{i+j} \ f_{i+j} \tag{7}$$

and we can exploit the results for the variation of the astrophysical factors presented in [14,13], see also [1].

First of all, as long as the *pp*-chain is the main source of solar energy, an increase of $S_{p+p}$ immediately implies a reduction of the central temperature $T_c$, in order that the solar luminosity ($L_\odot \propto S_{p+p} T^8$) is kept fixed. This implies:

$$T_c/T_c^{NOS} = (f_{p+p})^{-1/8} \tag{8}$$

where here and in the following the superscript NOS refers to the no-screening model. The temperature dependence of the neutrino fluxes is well known

$$\Phi_{Be}/\Phi_{Be}^{NOS} = (T_c/T_c^{NOS})^{10} \tag{9}$$

$$\Phi_{B}/\Phi_{B}^{NOS} = (T_c/T_c^{NOS})^{23.6} \tag{10}$$

$$\Phi_{N,O}/\Phi_{N,O}^{NOS} = (T_c/T_c^{NOS})^{22} \tag{11}$$

so that one can immediatly determine the relationship between the fluxes and the *pp*-enhancement factor.

Concerning the $^3He + ^3He$ and $^3He + ^4He$ reactions, we recall from ref. [14] that the $^7$Be-nuclei equilibrium concentration scales as:

$$N_{Be} \propto S_{^3He+^4He}/\sqrt{S_{^3He+^3He}} \tag{12}$$

thus the $^7$Be and $^8$B neutrino fluxes, which are both proportional to $N_{Be}$, depend on $f_{^3He+^4He}/\sqrt{f_{^3He+^3He}}$. Since the enhancement factors we consider are independent of the isotope, $f_{^3He+^4He} = f_{^3He+^3He} = f_{He+He}$ and thus $\Phi_{Be} \propto \sqrt{f_{He+He}}$. The $^8$B neutrinos, besides, depend linearly on the astrophysical factor $S_{pe+^7Be}$.

Regarding the CN neutrinos, we remind that the CN cycle is governed by the slowest reaction $^{14}N + p$, so that the CN-neutrino fluxes can be approximately considered as a linear function of $S_{p+^{14}N}$.

In conclusion, we can describe the relationship between the main components of neutrino fluxes and the enhancement factors by using the following equations:

$$\Phi_{Be} = \Phi_{Be}^{NOS} \ (f_{He+He})^{1/2} \ (f_{p+p})^{-10/8} \tag{13}$$

$$\Phi_B = \Phi_B^{NOS} \ (f_{He+He})^{1/2} \ f_{p+^7Be} \ (f_{p+p})^{-23.6/8} \qquad (14)$$

$$\Phi_{N,O} = \Phi_{N,O}^{NOS} \ f_{p+^{14}N} \ (f_{p+p})^{-22/8} \quad . \qquad (15)$$

The behaviour of *pp*-neutrinos can be derived best by using the conservation of luminosity, which implies (once minor neutrino flux components are neglected):

$$\Phi_{pp} + \Phi_{Be} + \Phi_N + \Phi_O = \Phi_{pp}^{NOS} + \Phi_B^{NOS} + \Phi_N^{NOS} + \Phi_O^{NOS}. \qquad (16)$$

The behaviour of *pep* neutrinos is derived by observing that their ratio to *pp* neutrinos is essentially constant in any solar model (see [15,16]):

$$\Phi_{pep}/\Phi_{pp} = \Phi_{pep}^{NOS}/\Phi_{pp}^{NOS} \quad . \qquad (17)$$

Note that in equations (13-15) the contributions of the enhancement factors corresponding to different reactions tend to compensate one with the other, so that the total variation of the flux is smaller than one would have if just one enhancement factor were introduced.

Note also that although $^3He+^3He$ and $^3He+^4He$ have the same enhancement factor, the equilibrium concentration of $^7Be$ nuclei, and thus $\Phi_{Be}$ and $\Phi_B$ are changed when screening is introduced. Concerning $\Phi_{Be}$, one has to remark the near cancellation between the $f_{p+p}$ and $f_{^3He+^3He}$ contributions.

By using the above equations with the enhancement factors given in Table II, we can quantitatively reproduce, to a large extent, the numerical results presented in section II, compare Table III and Table I.

## IV. A GENERALIZATION OF THE MITLER MODEL AND A MODEL INDEPENDENT ANALYSIS

None of the approaches to screening discussed above is completely satisfactory. The weak screening approximation is not justified for reactions other than the *pp*, since $Z_1 Z_2 e^2 \kappa/(kT)$ is not so small. The GDGC result stems from an interpolation of numerical computations and the prescription of the authors yields an unphysical discontinuity at the border between the weak and intermediate regimes. The CSK result, which incorporates the dynamical effects of a finite nuclear velocity, is anyhow derived in the framework of a linear theory, i.e. the weak screening approximation.

The Mitler result goes beyond the weak screening approximation, nevertheless the partial mobility of ions due to ions interaction effects and/or due to the finite thermal speed is not taken into account. Also, the value of the electron density at the nuclear site is somehow artificially kept equal to the average electron density $n_e$.

One can easily generalize Mitler formula by introducing two phenomenological parameters to overcome the above deficiences: we will keep eq. 4 but leave the ratio

$$n_e/n_o = \delta \qquad (18)$$

as a free parameter, in order to account for electron condensation around the nuclei. We also introduce an effective fraction of ions screening as:

$$n_{\text{eff}} = \gamma n_i \qquad (19)$$

where $\gamma$ is again a free parameter accounting for the partial mobility of the ions. Equation (2) is so replaced with

$$\kappa^2 = \frac{4\pi e^2}{kT}(\gamma \sum_i Z_i^2 n_i + n_e \theta_e) \quad . \qquad (20)$$

It is physically clear that both $\delta$ and $\gamma$ vary between 0 and 1.

In Fig. 2, we show the enhancement factors for the most relevant reactions in the Sun, in the plain $(\gamma, \delta)$. It is worth remarking several points: i) the very weak dependece on $\delta$, i.e. on the precise value on the electron density at nuclear site; ii) in the limit of small $\delta$ and $\gamma = 1$, i.e. large electron density at nuclear site and completely free ions, one recovers, obviously, the weak screening approximation, and this approximation - notwithstanding its name - gives the strongest effect; iii)the smallest value of the enhancement factors corresponds to the upper left corner ($\gamma = 0, \delta = 1$) since the ions do not contribute to the screening and the electron effect is as small as possible being no electron enhancement at nuclear site; iv) as $\delta$ is kept zero, by moving along the $\gamma$-axis, one explores dynamic ion effects in the weak screening regime, in particular the CSK model corresponds to $\gamma$ in the range $0.3 \div 0.5$; v) the enhancement factors following GDGC correspond to the limit of very small $\gamma$-s, which-we recall- is particulary insensitive to $n_o$. One thus sees that this simple phenomenological approach encompasses all the models discussed above (the original Mitler model clearly correspond to $\gamma = \delta = 1$)

Note that the enhancement factors for $He + He$ and $Be + p$ reactions, which are equal in the weak screening approximation, see eq. (1), are quite similar through all the plane.

In Fig. 3 we present the results for some neutrino fluxes ($^7$Be and $^8$B) and for the experimental signals in radiochemical experiments (Cl and Ga ) as a function of $\gamma$ and $\delta$, calculated by using the analytical expressions given in previous section. Again, the smallest values corresponds to the upper left corner ($\gamma = 0, \delta = 1$) and the largest ones to the lower right ($\gamma = 1, \delta = 0$), the Mitler original value being in between these two extrema. From inspection of the figure one concludes that deviations from the Mitler values are always smaller than 1% for the Beryllium flux and the Gallium signal, not exceeding 3% for the boron flux and the Chlorine signal. These conclusions are confirmed by explicit evaluations of solar models for the cases ($\gamma = 0, \delta = 1$) and ($\gamma = 1, \delta = 0$).

It is worth observing that even our smallest values are anyhow a few percent larger than those obtained by using the GDGC prescription. This latter has the largest enhancement factor for the $pp$-reaction; this implies the largest temperature reduction (with respect to NOS model), which tends to lower Beryllium and Boron neutrinos. This is partly compensated by the enhancement factors for the other reactions, which are the smallest ones and consequently both $\Phi_B$ and $\Phi_{Be}$ stay close to the no-screening values. In conclusion, this result originates from the unsatisfactory discontinuity between the weak and intermediate screening regime.

One can be even more general in addressing the role of screening effect on solar neutrinos. As clear from section III, the relevant parameters are $f_{p+p}, f_{He+He}, f_{p+^7Be}$ and $f_{p+^{14}N}$; furthermore we can take $f_{p+^7Be} = f_{He+He}$, as implied by the previous discussion. In this case one has

$$\Phi_{\text{Be}} = \Phi_{\text{Be}}^{\text{NOS}} \quad f_{\text{He+He}}^{1/2} \quad (f_{\text{p+p}})^{-10/8} \tag{21}$$

$$\Phi_{\text{B}} = \Phi_{\text{B}}^{\text{NOS}} \quad (f_{\text{He+He}})^{3/2} \quad (f_{\text{p+p}})^{-23.6/8} \tag{22}$$

In Fig. 4 we present the resulting $\Phi_{\text{Be}}$ and $\Phi_{\text{B}}$ obtained by varying $f_{\text{p+p}}$ in the range $1 \div 1.5$ and $f_{\text{He+He}}$ in the range $1 \div 6$ (so that the "screening energies", $U = kT \ln f$ at the upper estrema are 10 times larger than those in the original Mitler model). In the same figure we also plot the model independent information on Beryllium and Boron-neutrinos fluxes for standard neutrinos, derived directly by experimental data [17-20] for $\Phi_{\text{CNO}} = 0$ (this is the best possible case, for avoiding an unphysical conclusion $\Phi_{\text{Be}} < 0$, see [14,21]). One finds that by hugely varying $f_{\text{He+He}}$ and $f_{\text{p+p}}$ it is possible to reduce significantly the Boron-neutrino flux, but the Beryllium-neutrino flux remains anyhow larger than experimental data imply.

## V. CONCLUSIONS

We recall the main points of our discussion:
i) It seems to us that the Mitler model, equations (4,5), provides an essentially complete and consistent description of the screening effect in the Sun. Screening *à la Mitler* implies, with respect to the no-screening case, a temperature reduction of 0.5%, a 2% (8%) increase of Beryllium (Boron)-neutrinos flux and a 2% (12%) increase of the Gallium (Chlorine) signal.
ii) We investigated a generalization of the Mitler model, by introducing and varying two phenomenological parameters, and we conclude that uncertainties due to the screening effect are at the level of 1% for the Beryllium-neutrino flux and the Gallium signal, not exceeding 3% for the Boron-neutrino flux and the Chlorine signal. This means to us that the screening correction is well under control, and possible uncertainties are too small to be of significance for the solar neutrino problem.
iii) In a model independent way, where the enhancement factors for the $pp$ and $He + He$ reactions are kept as free parameters, allowing them to be even much larger than the standard Mitler values, we have calculated the resulting Beryllium and Boron neutrino flux and we have shown them to be essentially inconsistent with experimental data, see. Fig. 4.

## ACKNOWLEDGMENTS

We express our gratitude to V. Castellani for his useful comments, we also thanks M. Lissia for his advice. S. Degl'Innocenti acknowledges useful discussion with S. Turck-Chièze and H. Dzitko.

TABLES

TABLE I. Comparison among solar models with different screening predictions: NOS=no screening, WES=weak screening, MIT=Mitler 1977, GDGC=Graboske et al.1973 and CSK=Carraro et al.1988. We show the central temperature $T_c[10^{7o}K]$, the Helium abundance in mass Y, the metallicity fraction Z, the values of each component of the neutrino flux $[10^9 \text{cm}^{-2}\text{s}^{-1}]$, the calculated signals for the Chlorine (Cl) and the Gallium (Ga) experiments [SNU].

|  | NOS | WES | MIT | GDGC | CSK |
|---|---|---|---|---|---|
| $T_c$ | 1.573 | 1.566 | 1.566 | 1.564 | 1.567 |
| Y | 0.288 | 0.289 | 0.289 | 0.289 | 0.289 |
| Z ($\times 10^2$) | 1.85 | 1.85 | 1.84 | 1.84 | 1.85 |
| $pp$ | 60.0 | 59.6 | 59.7 | 60.0 | 59.7 |
| $pep$ | 0.146 | 0.142 | 0.142 | 0.143 | 0.143 |
| $^7$Be | 4.82 | 4.97 | 4.93 | 4.78 | 4.94 |
| $^8$B ($\times 10^3$) | 5.51 | 6.36 | 6.13 | 5.57 | 6.21 |
| $^{13}$N | 0.46 | 0.55 | 0.52 | 0.47 | 0.54 |
| $^{15}$O | 0.39 | 0.48 | 0.45 | 0.40 | 0.47 |
| Cl | 7.7 | 8.8 | 8.5 | 7.8 | 8.6 |
| Ga | 130 | 134 | 133 | 130 | 134 |

TABLE II. Enhancement factors for different screening prescriptions, calculated in central solar conditions. Same notations of Table I.

|  | WES | MIT | GDGC | CSK |
|---|---|---|---|---|
| $p+p$ | 1.049 | 1.045 | 1.049 | 1.038 |
| $He + He$ | 1.213 | 1.176 | 1.115 | 1.158 |
| $Be + p$ | 1.213 | 1.171 | 1.112 | 1.169 |
| $N + p$ | 1.403 | 1.293 | 1.192 | 1.324 |

TABLE III. Results of our analytical predictions, see text. Same notation as Table I.

|  | WES | MIT | GDGC | CSK |
|---|---|---|---|---|
| $T_c$ | 1.564 | 1.565 | 1.564 | 1.566 |
| $pp$ | 59.6 | 59.7 | 60.0 | 59.7 |
| $pep$ | 0.145 | 0.145 | 0.146 | 0.145 |
| $^7$Be | 5.0 | 4.95 | 4.80 | 4.96 |
| $^8$B ($\times 10^3$) | 6.39 | 6.15 | 5.61 | 6.21 |
| $^{13}$N | 0.57 | 0.53 | 0.48 | 0.55 |
| $^{15}$O | 0.48 | 0.45 | 0.41 | 0.47 |
| Cl | 8.8 | 8.5 | 7.9 | 8.6 |
| Ga | 135 | 133 | 130 | 134 |

# FIGURES

FIG. 1. For different reactions we plot the enhancement factors along the solar profile. We show the results of weak screening (dashed lines), Graboske et al. 1973 (dot-dashed lines), Mitler 1977 (straight lines) and Carraro et al. 1988 (dotted lines).

FIG. 2. For the most relevant reactions in the Sun, we show the value of the enhancement factors in a Mitler extended model, where the two parameters $\gamma$ and $\delta$ take into acount respectively the partial mobility of background ions and the electron density enhancement at nuclear site, see text. Labels refer to the solid curves and dashed curves are half-way the solid ones.

FIG. 3. We show the predicted valued of the ratios $\Phi_{Be}/\Phi_{Be}^{NOS}$, $\Phi_B/\Phi_B^{NOS}$ and the signals of radiochemical experiments (Chlorine and Gallium) in SNU, in an extended Mitler model as a function of the two parameters $\gamma$ and $\delta$, see text. Labels refer to the solid curves and dashed curves are half-way the solid ones.

FIG. 4. We present in the ($\Phi_{Be}$, $\Phi_B$) plane the regions corresponding to $f_{p+p}$ in the range $1 \div 1.5$ and $f_{He+He}$ in the range $1 \div 6$ (dashed area). We also report the regions allowed by the present experimental results, see ref. [21]. The full diamond show the values predicted by the no-screening model, the empty diamond the values predicted by the standard Mitler model.

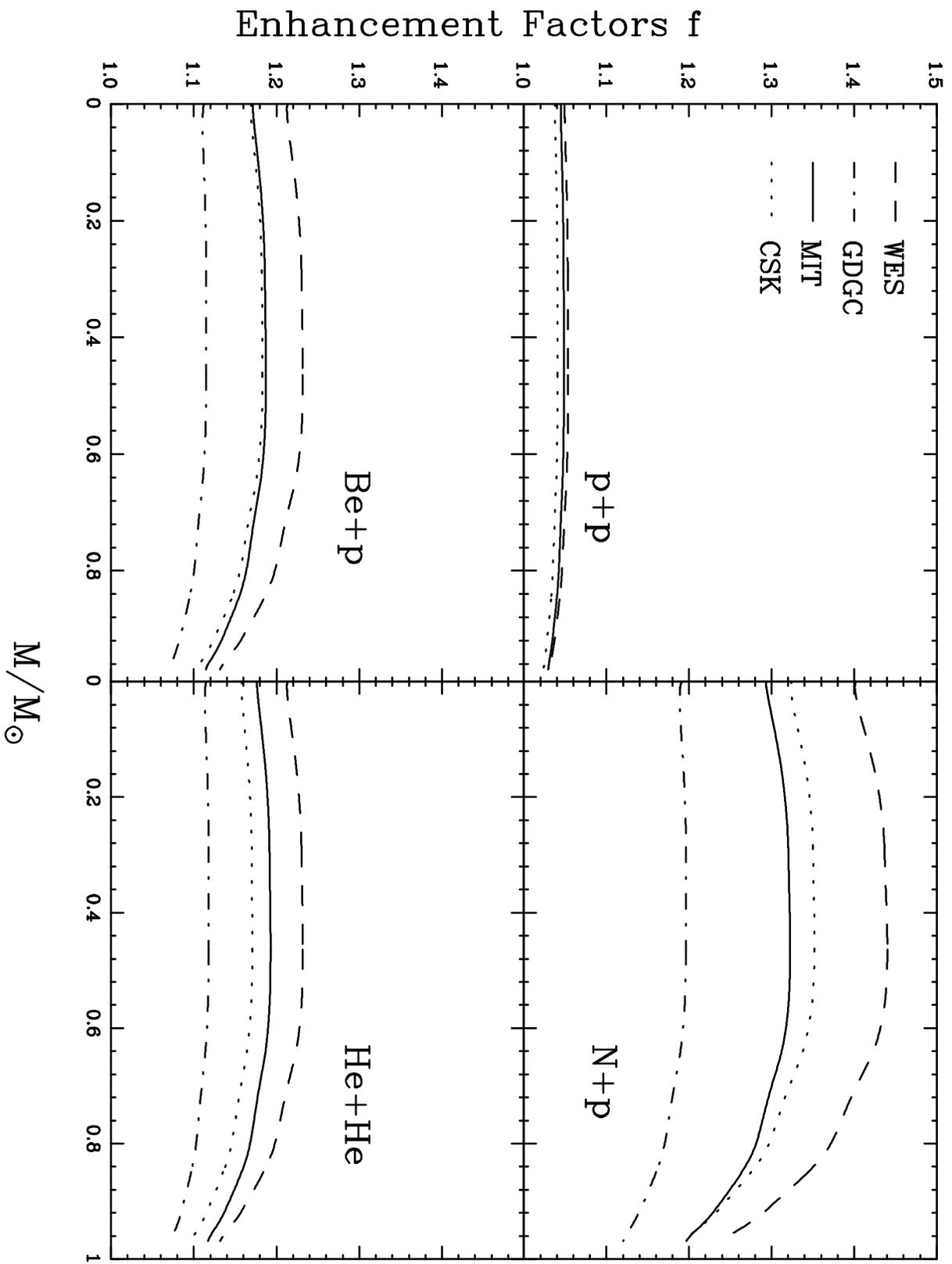

Fig. 1

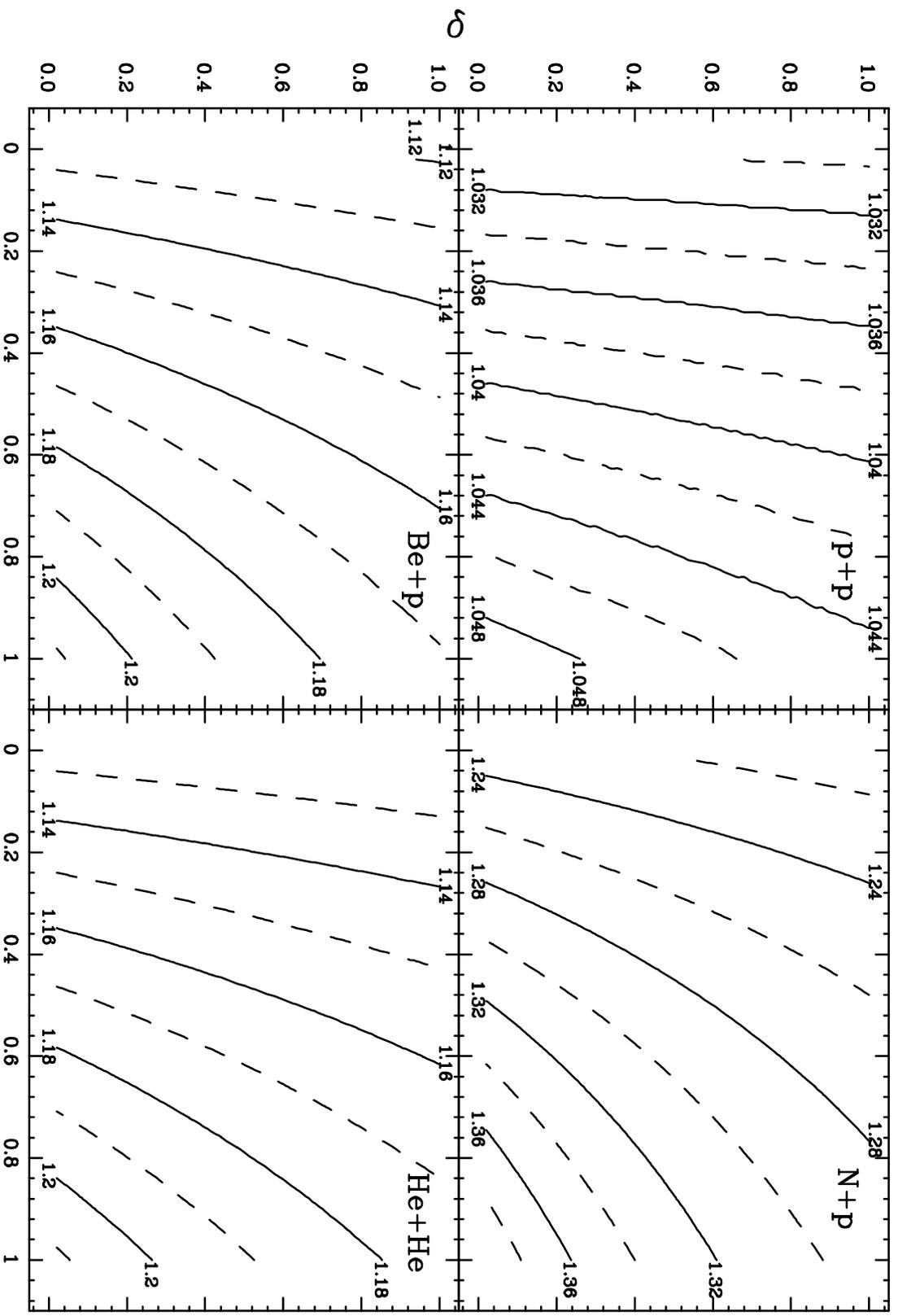

Fig. 2

Fig. 3

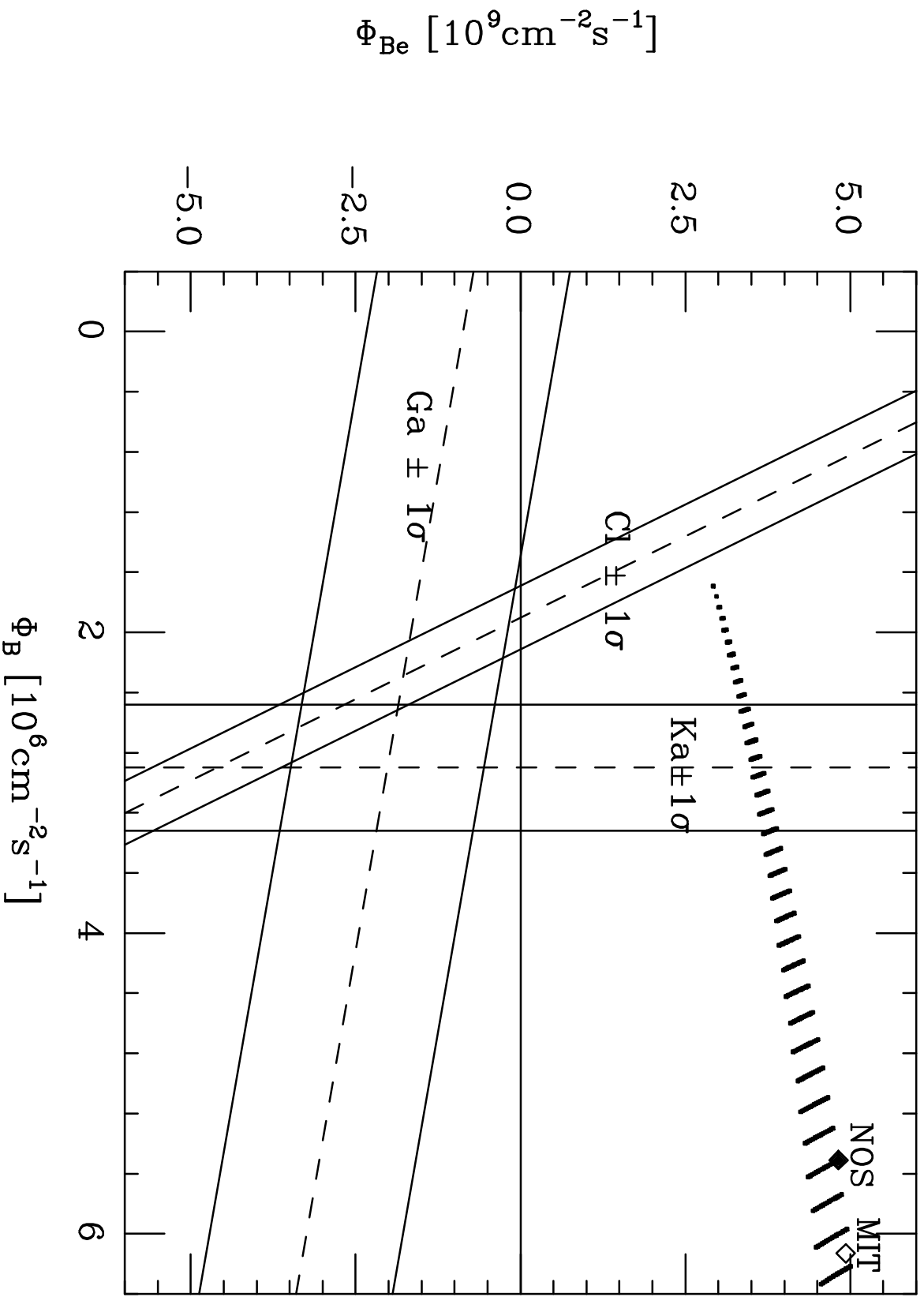

Fig. 4